\documentclass[prc,twocolumn,epsfig]{revtex4}
\usepackage{graphics}
\usepackage{epsfig}
\usepackage{amsfonts}
\usepackage{amsmath}
\usepackage{bm}

\usepackage{color}

\begin{document}
\title{Nuclear vorticity in isoscalar E1 modes: Skyrme-RPA analysis}
\author{P.-G. Reinhard$^1$, V.O. Nesterenko$^2$, A. Repko$^3$, and J. Kvasil$^3$}
\affiliation{$^1$ Institut f\"ur Theoretische Physik II,
Universit\"at Erlangen, D-91058, Erlangen, Germany}
\affiliation{$^2$ Laboratory of Theoretical Physics, Joint
Institute for Nuclear Research, Dubna, Moscow region, 141980,
Russia}
\affiliation{$^3$ Institute
of Particle and Nuclear Physics, Charles University, CZ-18000,
Prague 8, Czech Republic} 

\date{\today}

\begin{abstract}
Two basic concepts of  nuclear vorticity, hydrodynamical (HD) and
Rawenthall-Wambach (RW), are critically inspected. As a test case,
we consider the interplay of irrotational and vortical motion in
isoscalar electric dipole E1(T=0) modes in $^{208}$Pb, namely the
toroidal and compression modes. The modes are described in a
self-consistent random-phase-approximation (RPA) with the Skyrme
force SLy6. They are examined in  terms of strength functions,
transition densities, current fields, and formfactors. It is shown
that the RW conception (suggesting the upper component of the
nuclear current as the vorticity indicator) is not robust. The HD
vorticity is not easily applicable either because the definition
of a velocity field is too involved in nuclear systems. Instead,
the vorticity is better characterized by the toroidal strength
which closely corresponds to HD treatment and is approximately
decoupled from the continuity equation.

\end{abstract}


\maketitle

\section{Introduction}
\label{sec:introduction}

It is well known that the nuclear flow may be both irrotational
and vortical \cite{Bohr_book_74,Ring_book_80}. The irrotational
motion is presented by numerous examples of low-energy excitations
and electric giant resonances (GR) \cite{Harakeh_book_01} while
the vortical motion is exhibited by single-particle excitations
\cite{Ra87}, nuclear rotation \cite{Bohr_book_74}, and particular
GR (toroidal electric dipole \cite{Dub_75_83,Sem81} and twist
magnetic quadrupole \cite{HE_77,HE_79}).

Collective nuclear vorticity in electric GR is especially
interesting. Though  multipole electric GR are most irrotational,
there is a remarkable exception in the isoscalar E1(T=0) channel.
Here, after exclusion of the nuclear center-of-mass (c.m.) motion,
the vortical toroidal mode (TM) dominates in the low-energy (E$<$
10 MeV) E1(T=0) excitations \cite{Pa07,Kva_PRC_11}. So, in this
channel, the nuclear vorticity is realized as a leading mode. It
is remarkable that the TM lies in the energy region of so called
pygmy dipole resonance (PDR) and determines there the main flow
\cite{Ry02,Rep_PRC_13}. The low-energy strength (LES) in this
region is of high current interest as it can deliver useful
information on principle nuclear properties (nuclear symmetry
energy, neutron skin) with consequences to  various astrophysical
applications \cite{Pa07}. The vorticity can affect these relations
and thus deserves detailed analysis.

Despite some previous studies (see e.g.
\cite{Ra87,Sem81,Se83,Ca99}), our knowledge about the nuclear
vorticity is still poor. Even the basic points, the definition of
nuclear vorticity and choice of the proper observable, are
disputable. In hydrodynamics (HD), the vorticity is defined as a
curl of the velocity \cite{Lan87},
\begin{equation}\label{HDV}
  \vec \varpi ({\vec r})=
\vec{\nabla} \times \vec{v} (\vec{r}) \; .
\end{equation}
However, nuclear physics deals not with velocities but nuclear
currents. In this connection, Raventhall and Wambach have proposed
the $j_{+}(r)$-component of the nuclear current as an indicator of
the vorticity (RW vorticity in what follows) \cite{Ra87}. Indeed,
$j_{+}(r)$ may be posed as unrestricted by the continuity equation
(CE)
\begin{equation}\label{ce_ctd}
\delta\dot{\rho}_{\nu} (\vec{r}) + \vec{\nabla} \cdot \delta\vec{j}_{\nu}(\vec{r}) = 0
\end{equation}
(where $\delta\rho_{\nu}$ and $\delta\vec{j}_{\nu}$ are nucleon
and current transition densities for excited states $\nu$) and
thus suitable for a divergence-free (vortical) observable
\cite{Ra87,Ca99,Hei_PRC_82}. However, HD and RW definitions of the
vorticity strictly contradict each other \cite{Kva_PRC_11} when
being applied to the E1(T=0) compression mode (CM)
\cite{Har_PRL_77,Stri82}. Following HD, the CM velocity field is
\begin{equation}\label{HD_criterium}
  \vec{v}_{\text{CM}}(\vec{r})
  \propto
  \vec{\nabla}\, (r^{3} Y_{1\mu})
\end{equation}
and so this mode is fully irrotational. At the same time, the CM has
an essential $j_{+}(r)$-contribution \cite{Kva_PRC_11} and so, following RW, is
of a mixed (vortical/irrotational) character. This discrepancy certainly needs
a careful analysis.

The aim of the present paper is to scrutinize the HD and RW
prescriptions and finally propose the most relevant indicator and
measure of the nuclear vorticity. As shown below, the RW
prescription is not accurate and may result in wrong conclusions,
like in the CM case mentioned above. The HD prescription
(\ref{HDV}) is more physically transparent but not convenient for
practical use in nuclear physics. Instead, the toroidal strength
seems to be the most appropriate (though not perfect) measure of
nuclear vorticity in internal single-particle and collective
excitations. Toroidal strength can be considered as approximate HD
treatment in practicable form. It provides sufficiently good
decoupling from CE (\ref{ce_ctd}), avoids shortcomings of the RW
prescription, and exhibits a natural curl-like vortical motion.

Our analysis uses TM and CM as most relevant representatives of
the vortical and irrotational flows. Schematic images of these
modes in E1(T=0) channel are presented in Fig. 1. Note that the TM
and CM operators are related \cite{Kva_PRC_11}. Both modes
dominate the E1(T=0) channel and their maxima are well separated
in energy.

The calculations are performed for $^{208}$Pb within the Skyrme
random-phase-approximation (RPA) approach \cite{Rei92}. The method
is fully self-consistent in the sense that  both the mean field
and residual interaction are derived from the Skyrme functional
\cite{Skyrme,Vau72,En75,Ben03}. The residual interaction takes
into account all the terms of the Skyrme functional as well as the
Coulomb (direct and exchange) terms.  The Skyrme force SLy6
\cite{Sly6}, well describing the isovector (T=1) giant dipole
resonance (GDR) in heavy nuclei \cite{nest_PRC_08}, is used. This
study is a continuation of our previous exploration  of TM, CM,
and RW strengths \cite{Kva_PRC_11} within the self-consistent {\it
separable} random-phase-approximation (SRPA) method
\cite{nest_PRC_02,nest_PRC_06}. The present Skyrme RPA approach
does not use any separable approximation and implements a wider
configuration space.

A large set of dynamical characteristics is analyzed. Not only
strength functions but also current/velocity fields, and
formfactors are considered. What is important, curls and
divergences of TM and CM flows, as their natural signatures, are
inspected. These characteristics are refined from the dominant
Tassie collective modes (e.g. GDR or spurious center of mass
motion) whose curls and divergences are zero by definition. At the
same time, curls and divergences are effective fingerprints of
vortical TM and irrotational CM, which are not the Tassie modes.

The paper is organized as follows.  In Sec.~II, the basic
formalism for TM/CM modes and RW/HD vorticity prescriptions is
presented. In Sec.~III, the calculation scheme is outlined and the
dynamical characteristics to be explored (strength functions,
curent/velocity fields) are defined. In Sec. IV, the numerical
results are presented. In Sec.~V, various prescriptions of the
vorticity are discussed. The toroidal strength is shown to be the
most robust measure of the vorticity. In Sec.~VI, the summary is
done.

\section{Theoretical background}
\subsection{Basic expressions}

The standard electrical multipole operator reads \cite{BMv1}:
\begin{eqnarray}\label{ME_oper1}
\hat{M}(E\lambda\mu, k) &=&
-i \frac{(2 \lambda + 1)!!}{c k^{\lambda+1} (\lambda+1)}
\\
&&\cdot\int\!d^3r
\hat{\vec{j}}_{\text{nuc}}(\vec{r})\!\cdot\!
[\vec{\nabla}\!\times\!(\vec{r}
\!\times\!\vec{\nabla})j_{\lambda}(kr) Y_{\lambda \mu}(\hat{\vec r})]
\nonumber
\\
\nonumber
&=& \frac{(2 \lambda + 1)!!}{c k^{\lambda+1}}
\sqrt{\frac{\lambda}{\lambda+1}}
\\
&& \cdot\int d^3r \: [\: j_{\lambda}(kr)\:
\vec{Y}_{\lambda \lambda \mu}(\hat{\vec r}) \:]
\cdot [ \vec{\nabla} \times \hat{\vec{j}}_{\text{nuc}}(\vec{r})]
\nonumber
\end{eqnarray}
where $\hat{\vec{j}}_{\rm{nuc}}(\vec{r})=
\hat{\vec{j}}_{\rm{c}}(\vec{r})+\hat{\vec{j}}_{\rm{m}}(\vec{r})$
is operator of the nuclear current density consisting from the
convection and magnetization parts; $j_{\lambda}(kr)$ is the
spherical Bessel function; $\vec{Y}_{\lambda \lambda \pm 1
\mu}(\hat{\vec r})$ and $Y_{\lambda\mu} (\hat{\vec r})$ are vector
and ordinary spherical harmonics \cite{Va76}. Following
\cite{Kva_PRC_11}, the role of the magnetization current
$\hat{\vec{j}}_{\text m}(\vec{r})$  in E1(T=0) channel is
negligible. So only the convection  current $\hat{\vec{j}}_{\text
c}(\vec{r})$ is further involved. For the sake of brevity, we will
skip below (up to the cases of a possible confusion)  the
coordinate dependence in currents, densities and spherical
harmonics.
\begin{figure}[t]
\includegraphics[width=3.3cm,angle=-90]{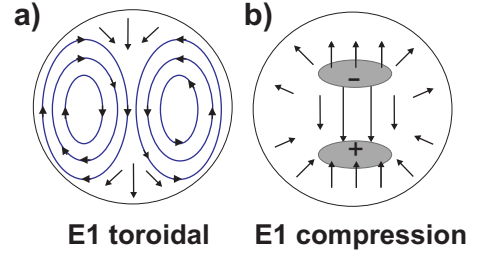}
\caption{(a) Schematic view of the E1(T=0)  toroidal (a) and
compression (b) modes \cite{Rep_PRC_13}. The driving field is
directed along z-axis. The arrows indicate the flow directions but
not the strengths. In the panel (b), the compression (+) and
decompression (--) regions, characterized by increased and
decreased density, are marked.}
\end{figure}

In the long-wave approximation ($k \to 0$), we get
\begin{equation}\label{E+ktor}
\hat{M}(E\lambda\mu, k) \approx \hat{M}(E\lambda\mu)
+ k\:\hat{M}_{\text{TM}}(E\lambda\mu)
\end{equation}
where
\begin{eqnarray}\label{E_oper_main}
\nonumber
\hat{M}(E\lambda \mu) &=&
 -\frac{i}{kc} \int d^3r \: (\vec{\nabla}
\cdot \hat{\vec{j}}_{\text c})
 r^{\lambda} Y_{\lambda \mu}
\\
 &=& - \int d^{3}r \:\hat{\rho}\:
 r^{\lambda} Y_{\lambda\mu}
\end{eqnarray}
is the familiar electric operator (with $\hat{\rho}$ being
the density operator) and
\begin{eqnarray}\label{tor_rel_1}
\hat{M}_{\text{TM}} (E\lambda \mu)
&=& \frac{i}{2c(\lambda+1)(2\lambda +3)}
\\
&&\cdot\int d^3r \hat{\vec{j}}_{\text{c}} \cdot [ \:\vec{\nabla}
\times \:(\vec{r} \times \vec{\nabla}) r^{\lambda +2} Y_{\lambda \mu}]
\nonumber
\\
\nonumber
  &=&
 -\frac{1}{2c}
 \sqrt{\frac{\lambda}{\lambda+1}}\,\frac{1}{2\lambda+3}
 \int d^3r \: r^{\lambda+2}\vec{Y}_{\lambda\lambda\mu}
\\ \label{TM_curlj}
 &&\cdot
  \left(\vec{\nabla} \times \hat{\vec j}_{\text c}\right)
\end{eqnarray}
is the toroidal operator \cite{Dub_75_83,Sem81,Kva_PRC_11,Kv03}.
This operator is the second order ($\sim k^2$) correction to the
dominant electric operator (\ref{E_oper_main}).  It becomes
dominant at $k\gg 0$.  Being determined by the curl
$\left(\vec{\nabla} \times \hat{\vec j}_{\text c}\right)$, the
toroidal flow is well (though not exactly, see discussion in Sec.
IV-D) decoupled from CE. For this reason the toroidal operator
cannot be presented through the nuclear density alone and needs
knowledge of the current distribution.

The CM operator reads \cite{Kva_PRC_11,Har_PRL_77,Stri82}
\begin{eqnarray}\label{CM_divj}
  \hat{M}_{\text{CM}}(E\lambda\mu)
  &=&
 -
\frac{i}{2c(2\lambda+3)}
 \int d^3r \: r^{\lambda+2}{Y}_{\lambda\mu}
  \left(\vec{\nabla} \cdot \hat{\vec j}_{\text{c}} \right)
\\
&=&
-k\frac{1}{2(2\lambda+3)}\int d^3r
\hat{\rho} r^{\lambda+2} Y_{\lambda\mu} \;
\\
&=&
-k \hat{M}'_{\text{CM}}(E\lambda\mu) \;,
\end{eqnarray}
where $\hat{M}'_{\text{CM}}(E\lambda\mu)$ is its familiar
density-dependent form \cite{Har_PRL_77,Stri82}. The CM operator
does not follow from the long-wave expansion of the initial
electric operator (\ref{ME_oper1}) but is introduced as a proper
probe operator for excitation of the isoscalar dipole giant
resonance \cite{Har_PRL_77,Stri82}. Unlike the TM case, this
operator may be presented in both current- and density-dependent
forms. As mentioned above, the velocity of the CM flow is a
gradient function, which justifies the irrotational (longitudinal)
character of the flow.

As was found in \cite{Kva_PRC_11}, the sum of the TM and CM operators gives
the operator responsible for RW vorticity:
\begin{equation}\label{UM=TM+CM}
\hat{M}_{\text{RW}}(E\lambda\mu)= \hat{M}_{\text{TM}}(E\lambda\mu)
 + \hat{M}_{\text{CM}}(E\lambda\mu) \;.
\end{equation}
This relation makes possible a direct comparison of RW, TM, and CM
strengths. Besides, it shows that all these three operators are of
the second order with respect to the electric operator
(\ref{E_oper_main}).

\subsection{E1(T=0) case}

In E1(T=0) channel, the RW, TM, and CM operators are reduced to
\begin{eqnarray}
\label{RW_oper}
&& \hat M_{\text{RW}}(E1\mu) = -\frac{i}{5c}
\sqrt{\frac{3}{2}} \int d^3r  \hat{\vec{j}}_{c} r^{2} \vec Y_{12 \mu} ,
\\
\label{TM_oper} && \hat M_{\text{TM}}(E1\mu) = -\frac{i}{2\sqrt{3}c} \int
d^3r \hat{\vec j}_{c}
\\ \nonumber
&&\quad  \cdot \; [\frac{\sqrt{2}}{5}r^2\vec Y_{12\mu} + (r^2
-\langle r^2\rangle_0) \vec Y_{10\mu}] ,
\\
\label{CM_oper}
&& \hat M_{\text{CM}}(E1\mu) = -\frac{i}{2\sqrt{3}c} \int
d^3r \hat{\vec j}_{c}
\\ \nonumber
&&\quad  \cdot \; [\frac{2\sqrt{2}}{5} r^2\vec Y_{12\mu} - (r^2 -
\langle r^2\rangle_0) \vec Y_{10\mu}]
\\
\label{CMp_oper} && \hat M'_{\text{CM}}(E1\mu) =  \frac{1}{10}
\int d^3r \hat{\rho} \; [r^3-\frac{5}{3}\langle r^2\rangle_0 r]
Y_{1\mu} \; .
\end{eqnarray}
Here, $\langle r^2 \rangle_0^{\mbox{}}=\int d^3r \rho_0 r^2/A$ is
the ground-state squared radius, $\rho_0({\vec r})$ is the ground
state density, $A$ is the mass number. The operators
(\ref{TM_oper})-(\ref{CMp_oper}) have  the center of mass
correction (c.m.c.) proportional to  $\langle r^2\rangle_0$, while
in (\ref{RW_oper}) the c.m.c. is zero. In what follows, we
consider only $\mu$=0 case and thus skip the $\mu$ index.

The RW, TM, and CM matrix elements for $E1$ transitions between
the ground state $|0\rangle$ and RPA excited state $|\nu\rangle$
can be determined through the current transition density
\begin{equation}\label{CTD}
\delta \vec{j}^{\nu}(\vec{r}) = \langle \nu| \:
\hat{\vec{j}}_{\text c}(\vec{r}) \:| 0 \rangle = [j^{\nu}_{10}(r)
\:\vec{Y}_{10}^* + j^{\nu}_{12}(r) \:\vec{Y}_{12}^*]
\end{equation}
as
\begin{equation}
\label{RW_me}
 \langle \nu |\hat M_{\text{RW}}(E1)|0\rangle = -\frac{1}{5\sqrt{2}c}
\int dr  r^{4} j^{\nu}_{12} \; ,
\end{equation}
\begin{eqnarray}
\label{TM_me}
 \langle \nu |\hat M_{\text{TM}}(E1)|0\rangle &=& -\frac{1}{6c} \int
dr r^2
\\ \nonumber
&\cdot&  [\frac{\sqrt{2}}{5}r^2 j_{12}^{\nu}
+ (r^2 -\langle r^2\rangle_0) j_{10}^{\nu}] \; ,
\\
\label{CM_me}
 \langle \nu |\hat M_{\text{CM}}(E1)|0\rangle &=& -\frac{1}{6c} \int
dr r^2
\\ \nonumber
&\cdot&  [\frac{2\sqrt{2}}{5}r^2 j_{12}^{\nu}
- (r^2 -\langle r^2\rangle_0) j_{10}^{\nu}] \; .
\end{eqnarray}
The upper and lower current components are usually denoted as
$j_+$ and $j_-$ ($j_+=j_{12}$ and $j_-=j_{10}$ in E1 case). In
accordance to \cite{Ra87}, just $j_+$ determines the vorticity
(and RW matrix element (\ref{RW_me})). The flow can be fully
vortical ($j_+\ne$0, $j_-$=0), fully irrotational ($j_+=$0,
$j_-\ne$0), and mixed ($j_+\ne$0, $j_-\ne$0). Following this
prescription, both TM and CM are of a mixed
(irrotational/vortical) character, which contradicts with
predominantly  curl- and gradient-like velocities of these flows
\cite{Kva_PRC_11}.

\subsection{Hydrodynamical vorticity}

To analyze the HD vorticity (\ref{HDV}), we should define the velocity of
a nuclear motion and build the corresponding matrix elements. This can be
done by definition of the velocity transition density through the current one
\cite{Se83},
\begin{equation}\label{HDv}
\delta \vec{v}^{\nu}(\vec{r})=   \frac{\delta \vec{j}^{\nu} (\vec r)}{\rho_0(\vec r)}
\;,
\end{equation}
and the replacement
\begin{equation}\label{repl}
 \left[ \vec{\nabla} \times \delta {\vec j}^{\nu}(\vec r) \right]
 \to
\rho_0 (\vec{r}) \left[ \vec{\nabla} \times \delta {\vec v}^{\nu}(\vec r)\right]
\end{equation}
in the relevant matrix elements. It is easy to see from the exact
expression
\begin{equation}\label{replc}
 \vec{\nabla} \times \delta {\vec j}^{\nu}(\vec r)
 =
\rho_0 (\vec{r}) \vec{\nabla} \times \delta {\vec
v}^{\nu}(\vec r) + \vec{\nabla} \rho_0 (\vec{r}) \times
\delta {\vec v}^{\nu}(\vec r)
\end{equation}
that the replacement (\ref{repl}) neglects
$\vec{\nabla} \rho_0 (\vec{r})$ and thus a large change of
$\rho_0 (\vec{r})$ at the nuclear surface. So, the HD vorticity
build from (\ref{repl}) is relevant by construction only
at nuclear interior.

Among RW, TM, and CM operators, only the TM one (\ref{TM_curlj})
and its matrix element
\begin{eqnarray}
\label{TM_curl_me} &&  \langle \nu |\hat
M_{\text{TM}}(E1)|0\rangle = -\frac{1}{10\sqrt{2}c} \int d^3r
\\
\nonumber
&\cdot&  [r^3-\frac{5}{3}r \langle r^2\rangle_0] \; \vec Y_{11}
\cdot [\vec{\nabla} \times \delta \hat{\vec j}^{\nu}]
\end{eqnarray}
have the necessary curl-of-current structure suitable for using
the replacement  (\ref{repl}). Then, by substituting  (\ref{repl})
to (\ref{TM_curl_me}), we get the matrix element
\begin{eqnarray}
\label{HDV_me} &&  \langle \nu |\hat M_{\text{HD}}(E1)|0\rangle =
-\frac{1}{10\sqrt{2}c} \int d^3r
\\
\nonumber
&\cdot& [r^3-\frac{5}{3}r \langle r^2\rangle_0] \; \rho_0 \vec Y_{11}
\cdot  [\vec{\nabla} \times \delta \hat{\vec v}^{\nu}]
\; ,
\end{eqnarray}
characterizing the HD vorticity.  The explicit expressions for
curls and divergences of $\delta \hat{\vec j}^{\nu}$ and $\delta
\hat{\vec v}^{\nu}$ are given in the Appendix \ref{Sec:app1}.

Note that, though the general electric operator (\ref{ME_oper1}) also has the
curl-of-current term, it cannot be used for building HD matrix elements
through the replacement (\ref{repl}). Indeed the vorticity is the second-order
divergence-free effect vanishing in the long-wave ($k \to 0$) approximation (LWA).
Instead, the operator (\ref{ME_oper1}) still has the LWA contribution.

Definition of the velocity (\ref{HDv}) has a well known
shortcoming. Being inverse to the density $\rho_0(\vec r)$, the
velocity becomes artificially large at the nuclear surface and
beyond. This shortcoming persists in the HD matrix elements
(\ref{HDV_me}). As was mentioned above in connection to Eq.
(\ref{replc}), the HD vorticity build from (\ref{repl})  can
anyway be applied only to the nuclear interior where $\rho_0(\vec
r)$ changes smoothly. The toroidal flow is similar to HD in the
interior but has  a good behavior at the surface. Thus the TM
strength is a more robust measure of the HD vorticity than the
construction (\ref{HDV_me}). This will be confirmed in Sec. IV by
numerical results.

\section{Method}

For analysis of the nuclear vorticity, a representative set of
variables is used: strength functions, flow patterns and
coordinate-energy maps for current (velocity) transition densities
and their derivatives (curls and divergences), and form-factors.

\subsection{Strength function}

The energy distribution of the mode strengths is described by the strength
function
\begin{equation}
\label{eq:strength_function}
  S_{\alpha}(E1; \omega) = 3
\sum_{\nu} \omega_{\nu}^l
  |\langle\nu|\hat{M}_{\alpha}(E1)|0\rangle|^2
  \zeta(\omega - \omega_{\nu})
\end{equation}
involving the Lorentz weight
\begin{equation}
  \zeta(\omega - \omega_{\nu}) =
  \frac{1}{2\pi}\frac{\Delta}{(\omega- \omega_{\nu})^2+\frac{\Delta^2}{4}}
\end{equation}
with the smoothing width $\Delta$. The type of the transition
operator $\hat{M}_{\alpha}(E1)$ is determined by the index $\alpha
= \{\text{E1, RW, TM, CM, HD}\}$, $\nu$ runs over the RPA spectrum
with eigen-frequencies $\omega_{\nu}$ and eigen-states
$|\nu\rangle$. The E1(T=1) strength function ($\alpha = E1$) uses
the energy weight ($l=1) $ and the ordinary E1 operator with the
effective charges $e_{\text{eff}}^n=-Z/A$ and
$e_{\text{eff}}^p=N/A$ (see the operator $\hat{D}_1$ below). Other
strength functions with $\alpha = \{\text{RW, TM, CM, HD}\}$ skip
the energy weight ($l=0$) and, being studied in T=0 channel, use
$e_{\text{eff}}^n=e_{\text{eff}}^p=1$.

\subsection{Flow patterns and coordinate-energy maps}

The strength functions provide a first overview of the modes. A
more insight can be gained by inspection of the current (velocity)
transition densities and their derivatives.

Since we are interested in general features of the modes, it is
convenient to consider the {\it integral} variables (involving
contributions from all the RPA states in a given energy interval $[E_1,E_2]$)
\begin{eqnarray}
\label{int_tcd_A}
  \vec{A}^{(D)}(\vec{r}) &=& \sum_{\nu \epsilon [E_1, E_2]}
  D_{\nu}^* \vec{A}^{\nu}(\vec{r}) \; ,
\\
\label{int_tcd_B}
  B^{(D)}(\vec{r}) &=& \sum_{\nu \epsilon [E_1, E_2]}
  D_{\nu}^* B^{\nu}(\vec{r})
\end{eqnarray}
or {\it average} variables (smoothed by the Lorentz weight
$\zeta$)
\begin{equation}
\label{aver_tcd}
 C^{(D)}(r, \omega)= \sum_{\nu}
  D_{\nu}^* C^{\nu}(r) \zeta(\omega-\omega_{\nu}) \; .
\end{equation}
The vectors variables $\vec{A}^{(D)}(\vec{r})$ give the flow
patterns describing in detail the coordinate (radial and angular)
distribution of the modes. The vector contributions
$\vec{A}^{\nu}(\vec{r})$ could be the curent/velocity transition
densities, their components and curls. Further, the variables
$B^{(D)}(\vec{r})$ provide the similar coordinate distribution but
for the scalar patterns like divergences of the flows. The
coordinate-energy maps $C^{(D)}(r,\omega)$ deliver information on
radial/energy distribution, thus combining properties of the
transition densities and strength functions. Using
(\ref{int_tcd_A})-(\ref{aver_tcd}) allows to avoid individual
details of RPA states but highlight their common features.

The calculation of such variables needs a precaution because of
arbitrary signs of RPA $\nu$-states. To overcome this trouble, we
use the technique \cite{Rep_PRC_13} where the values of interest
are additionally weighted by the matrix elements $D_{\nu}=\langle
\nu|\hat{D}_T(E1)|0\rangle$ of the dipole probe operator
$\hat{D}_T(E1)$. Then every state $\nu$ contributes to
(\ref{int_tcd_A})-(\ref{aver_tcd}) twice and thus the ambiguity is
removed. Two dipole probe operators are implemented: isovector
\begin{equation}
\hat{D}_1(E1)=(N/A)\sum_i^Z (r Y_{1})_i
-(Z/A)\sum_i^N (r Y_{1})_i
\end{equation}
for the GDR strength ($\alpha = E1$) and isoscalar
\begin{equation}
\hat{D}_0(E1)=\sum_i^A (r^3 Y_{1})_i
\end{equation}
for the modes $\alpha = \{\text{RW, TM, CM, HD}\}$.

For example, for the current transition density $\delta \vec{j}(\vec{r})$
and its radial component ${j}_{21}(r)$, the variables (\ref{int_tcd_A})
and (\ref{aver_tcd}) read
\begin{eqnarray}
\label{int_dj}
  \delta\vec{j}^{(D_0)}_{1}(\vec{r}) &=& \sum_{\nu \epsilon [E_1, E_2]}
  D_{\nu}^* \delta\vec{j}^{\nu}_{1}(\vec{r}) \; ,
\\
\label{aver_dj21}
 \delta j^{(D_0)}_{12}(r, \omega) &=& \sum_{\nu}
  D_{\nu}^* \delta j^{\nu}_{12}(r) \zeta(\omega-\omega_{\nu}) \; .
\end{eqnarray}
The explicit expressions for other cases are given in the Appendix \ref{Sec:app2}.

\subsection{Form-factors}

The form-factors are obtained from the average variables
(\ref{aver_tcd}) by the Fourier-Bessel transformation
\begin{equation}\label{ff}
 F^{(D)}(k, \omega) =  \sum_{\nu}
  D_{\nu}^* \zeta(\omega-\omega_{\nu}) \int dr r^2 j_1{(kr)} B^{\nu}(r)
\end{equation}
where $j_1{(r)}$ is the dipole spherical Bessel function.

\subsection{Calculation details}

The calculations are performed within the one-dimensional (1D)
Skyrme RPA approach \cite{Rei92,Ben03}. The approach is fully
self-consistent in the sense that  both the mean field and
residual interaction are derived from the Skyrme functional
\cite{Skyrme,Vau72,En75,Ben03}. Besides the residual interaction
takes into account all terms of the Skrme functional as well as
the Coulomb (direct and exchange) terms. There is no variational
c.m.c. term in the functional. The calculations are performed for
the doubly-magic nucleus $^{208}$Pb. We use the Skyrme force SLy6
\cite{Sly6} which provides a satisfactory description of the giant
dipole resonance (GDR) in heavy nuclei \cite{nest_PRC_08}.

The calculations employ a 1D spherical coordinate-space grid with
the mesh size 0.3 fm and a calculational box of 21 fm. A large RPA
expansion basis is used. The particle-hole ($1ph$) states are
included up to an excitation energy of $\sim 35$ MeV. Furthermore,
we employ a couple of fluid dynamical basis modes \cite{Rei92},
which allows to: i) include global polarization effects up to 200
MeV, ii) provide correct extraction of the center-of-mass mode,
and iii) produce 100\% exhaustion of the energy-weighted sum rules
for isovector \cite{Ring_book_80} and isoscalar
\cite{Harakeh_book_01} GDR.

\section{Numerical results}
\subsection{Strength functions}

\begin{figure}
\includegraphics[width=7cm]{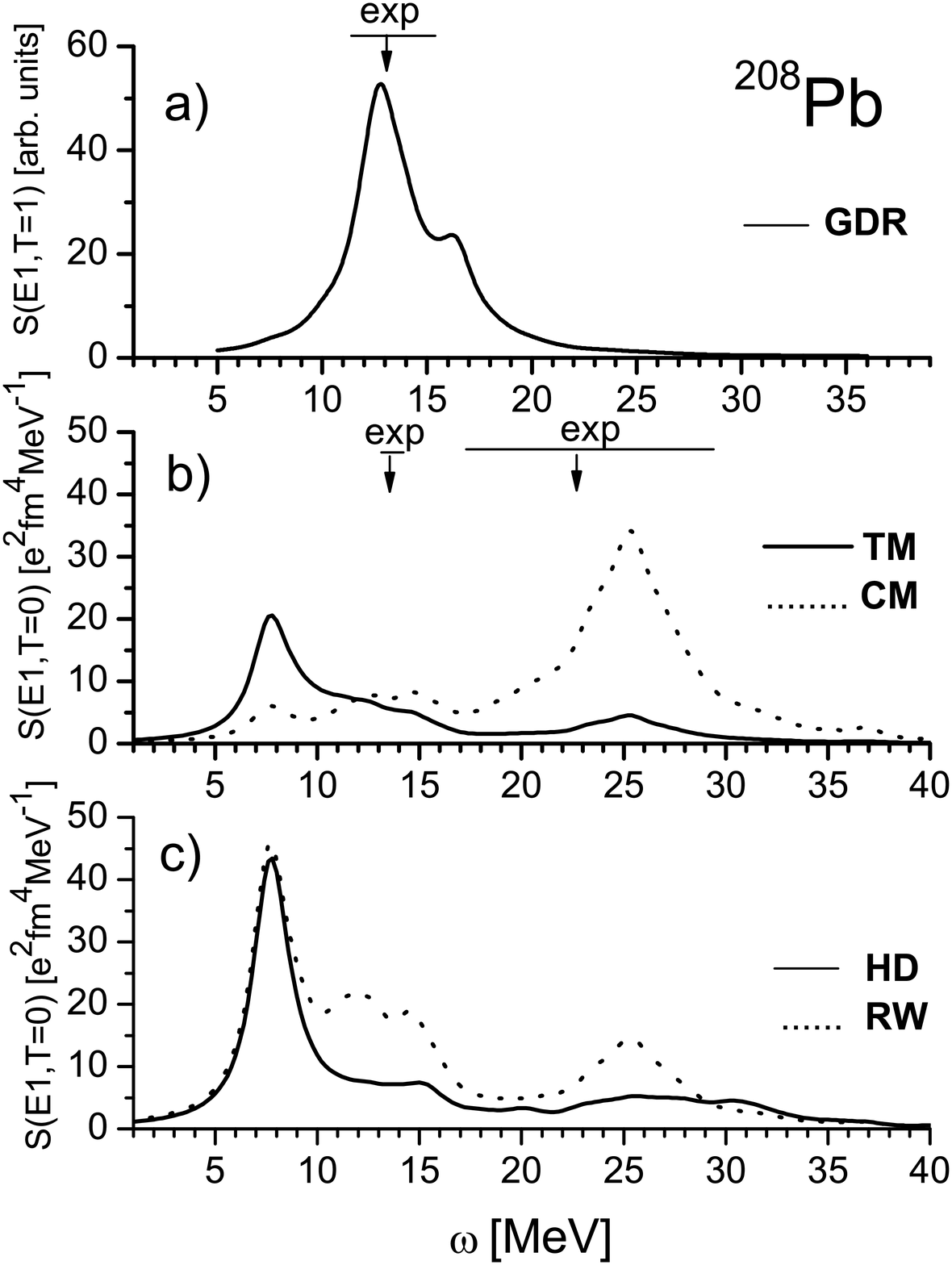}
\caption{The RPA strength functions. (a) E1(T=1) giant resonance.
The line with the arrow indicate the experimental width and energy
centroid of the resonance \cite{GDR_exp}. (b) E1(T=0) toroidal
(TM, solid line) and compression (CM, dotted line) strengths. The
widths and energy centroids of the low- and high-energy branches
of E1(T=0) excitations observed in ($\alpha, \alpha'$) reaction
\cite{Uchida_PLB_03,Uchida_PRC_04} are denoted. c) The E1(T=0)
hydrodynamical (HD, solid line) and Rawenthal-Wambach   (RW,
dotted line) vortical strengths.}
\end{figure}

In Fig. 2, some RPA strength functions in $^{208}$Pb are
exhibited. In panel (a), the calculated isovector GDR is compared
to the experimental data \cite{GDR_exp}. A good agreement with the
experiment justifies a satisfactory accuracy of our description.

Further, panels (b,c) demonstrate the TM, CM, HD, and RW strengths
in E1(T=0) channel. Note that, due to the large configuration
space and c.m.c. in the transition matrix elements (\ref{TM_me}),
(\ref{CM_me}), and (\ref{HDV_me}), the spurious strength is fully
downshifted below 0.5 MeV and thus does not affect the results.
The panel (b) shows that the calculated TM and CM strengths are
peaked at 7-8 MeV and $\sim$25 MeV, respectively. These results
somewhat deviate from available experimental $(\alpha,\alpha')$
data for E1(T=0) resonance \cite{Uchida_PLB_03,Uchida_PRC_04}
which give maxima  at 12.7 and 23.0 MeV. Such discrepancy is
common for various theoretical approaches \cite{Pa07} and worth to
be commented in more detail.

First of all, following the panel (b), the measured E1(T=0)
resonance may be treated as manifestation of the CM alone, i.e.
without TM contribution. Indeed, the experimental peaks at 12.7
and 23.0 MeV can correspond to the CM structures at 13-15 and 25
MeV in our RPA calculations. The familiar interpretation of the
experimental peak at 12.7 MeV as TM
\cite{Uchida_PLB_03,Uchida_PRC_04} is questionable since the
calculated TM lies much lower at 7-8 MeV. The experiment
\cite{Uchida_PLB_03,Uchida_PRC_04} explores the excitation energy
interval 8-35 MeV and thus perhaps loses the strong and narrow TM
peak at 7-8 MeV. Moreover, the $(\alpha,\alpha')$ reaction, being
mainly peripheral, is generally not suitable for observation of
the vortical TM.

Further, the discrepancy for CM energy (25 MeV in the theory
versus 23 MeV in the experiment) may be explained by a sensitivity
of this high-energy strength to the calculation scheme, in
particular to the size of the configuration space. The larger the
space, the lower the CM energy. It seems that even our impressive
space size (up to $\sim$200 MeV)  is not yet enough. Perhaps, the
coupling to complex configurations has here some effect.

Our RPA results are close to the previous relativistic
\cite{Vr00,Vr02} and Skyrme nonrelativistic \cite{Co00,Kva_PRC_11}
studies, including SRPA ones \cite{Kva_PRC_11}. The TM lies at 6-9
MeV, i.e. in PDR location. Following \cite{Rep_PRC_13}, the
E1(T=0) strength in this region has a complex composition with a
strong toroidal fraction.

Figure 2(c) exhibits the RW and HD strengths calculated with the
transition dipole matrix elements (\ref{RW_me}) and
(\ref{HDV_me}), respectively. As mentioned above, both them were
proposed as the vortical fingerprints. It is seen that RW and HD
give about equal strong peaks at 7-8 MeV, i.e. just at the TM
energy. So both them signal on the truly TM vortical motion.
However, the RW and HD  deviate at higher energies. The HD, being
similar to TM by construction, is modest everywhere with exception
of the TM region. Instead, the RW has additional maxima at the GDR
(10-15 MeV) and CM (25 MeV) regions, characterized by strong
irrotational flows. This means that RW is not a robust measure of
the vorticity.
\begin{figure}
\includegraphics[width=9cm]{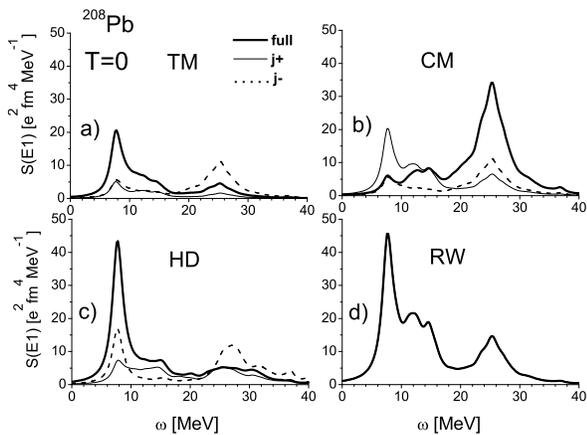}
 \label{fig:fig4}
\caption{(a) Toroidal (TM), (b) compression (CM), (c)
hydrodynamical (HD), and (d) Rawenthal-Wambach (RW) RPA strength
functions for the full nuclear current (bold line) and its
$j_-\equiv j_{10}$ (dotted line) and $j_+ \equiv j_{12}$ (thin
line) components. For RW case, $j_-=0$ and so only the $j=j_+$
strength is shown.}
\end{figure}

In Fig.~3, the contributions of $j_-(r) \equiv j_{10}(r)$ and
$j_+(r)  \equiv j_{12}(r)$ components of the nuclear current
(\ref{CTD}) to E1(T=0) TM, CM, RW, and HD strengths are
demonstrated. It is seen that both components are peaked in
low-energy (LE)  and high-energy (HE) regions, with some
preference of LE  for $j_+$ and HE for $j_-$. Following
expressions (\ref{TM_me}), (\ref{CM_me}), (\ref{HDV_me}), and
Appendix A, the TM, CM, and HD strengths are produced by
constructive or destructive interference of $j_+$ and $j_-$ (or
$v_+$ and $v_-$) contributions. The LE interference is
constructive for TD/HD and destructive for CM. For HE, the picture
is opposite. The RW  is by construction fully determined by $j_+$.
There is no seen any essential advantage of $j_+$ over $j_-$ to
represent the nuclear vorticity. Both components are almost
equally active in the vortical TM at 7-8 MeV and irrotational CM
at 25 MeV. This once more distrusts $j_+$ as a vortical
descriptor.
\begin{figure}
\includegraphics[width=7.5cm]{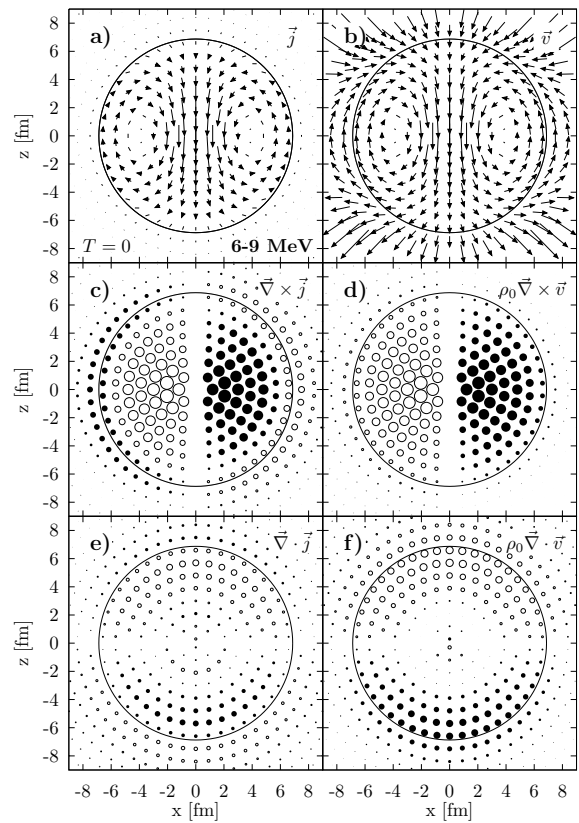}
 \label{fig:fig4}
\caption{E1(T-0) current and velocity flow patterns
(\ref{int_tcd_A})-(\ref{int_tcd_B}) at x-z plane for the TM energy
bin 6-9 MeV in $^{208}$Pb.:(a) nuclear current $\vec j$, (b)
nuclear velocity $\vec v$, (c) current curl $\vec{\nabla} \times
\vec j$, (d) velocity curl $\rho_0 \vec{\nabla} \times \vec v$, e)
current divergence $\vec{\nabla} \cdot \vec j$, and f) velocity
divergence  $\rho_0 \vec{\nabla} \cdot \vec v$. In the panels
(c)-(f), the flows, being perpendicular to the plane x-z, are
exhibited by circles. The open (filled) circles represent flows
along (opposite to) the y-axis. The flow magnitude is depicted by
the size of the arrows and circles (in arbitrary units). The
nuclear boarder is marked by a large circle. See explicit
expressions for patterns in  Appendix B.}
\end{figure}

\subsection{Flow patterns}

As compared to the strength functions, the flow patterns  deliver
a more detailed information on nuclear dynamics. Here we depict
not only nuclear current and velocity fields but also their
divergences and curls. The patterns $\vec{\nabla}\cdot \vec{j}$
and $\vec{\nabla}\times \vec{j}$ are especially important since
they directly indicate if the current contributes to CE.
Obviously, only curl-free ($\vec{\nabla}\times \vec{j}$=0)
currents are irrotational and coupled to CE. Instead, the
divergence-free ($\vec{\nabla}\cdot \vec{j}$=0) currents carry the
vorticity and are CE-unrestricted.

Note that the isovector GDR and isoscalar spurious c.m. motion are
basically driven by the operator $r Y_{1\mu}$ with the velocity
field $\vec{v} \propto \vec\nabla (r Y_{1\mu})$. They are the
collective Tassie modes with $\vec{\nabla}\cdot
\vec{j}=\vec{\nabla}\times \vec{j}=0$ and thus do not contribute
to the curl and divergence patterns. Instead, the E1 TM and CM are
characterized by the operators with $r^3$-dependence and so do not
belong the Tassie modes. For them, the patterns $\vec{\nabla}\cdot
\vec{j}$ and $\vec{\nabla}\times \vec{j}$ become indeed
informative.
\begin{figure}
\includegraphics[width=8cm]{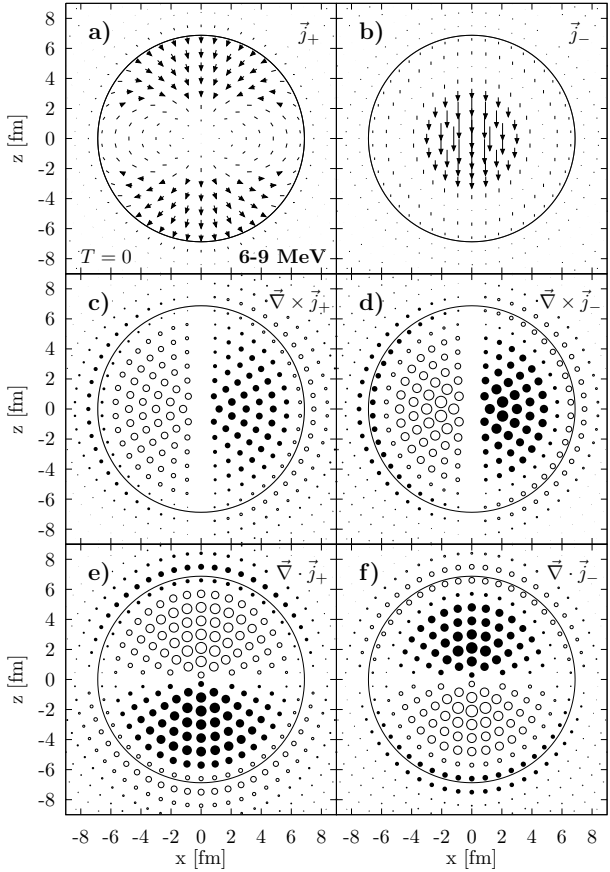}
 \label{fig:fig4}
\caption{E1(T=0) flow patterns for the current components $\vec
j_+$ (left) and $\vec j_-$ (right) at x-z plane for the TM energy
bin 6-9 MeV in $^{208}$Pb. The current components $\vec j_{\pm}$
(a)-(b), as well as their curls (c)-(d) and divergences (e)-(f)
are exhibited. Like in Fig. 4, the presentation is in terms of
arrows and circles. See explicit expressions for the patterns in
Appendix B.}
\end{figure}

In Fig. 4, different patterns for the energy bin 6-9 MeV
containing the TM are considered. The panel (a) shows that, in
accordance with our previous study \cite{Rep_PRC_13}, the current
$\vec{j}$ is mainly of the toroidal nature (compare to Fig 1(a)
for the schematic image of TM). The same takes place for the
velocity field $\vec{v}$ exhibited in the panel (b). The velocity
is not damped by the density factor and so is artificially strong
at the nuclear surface (marked by the circle of the radius $R$ =
1.16 fm $A^{-1/3}$) and beyond. Following panels (c) and (e), the
current curl is much stronger than its divergence, which confirms
basically vortical character of the flow. The density-weighted
curl and divergence of the velocity (panels (d)-(f)) are very
similar to their current counterparts in the nuclear interior. A
difference takes place only at the nuclear surface. So, up to the
surface region,  the HD vorticity determined by
$\vec{\nabla}\times \vec{v}$ can be well characterized by
$\vec{\nabla} \times \vec j$.
\begin{figure} \label{fig:fig6}
\includegraphics[width=8cm]{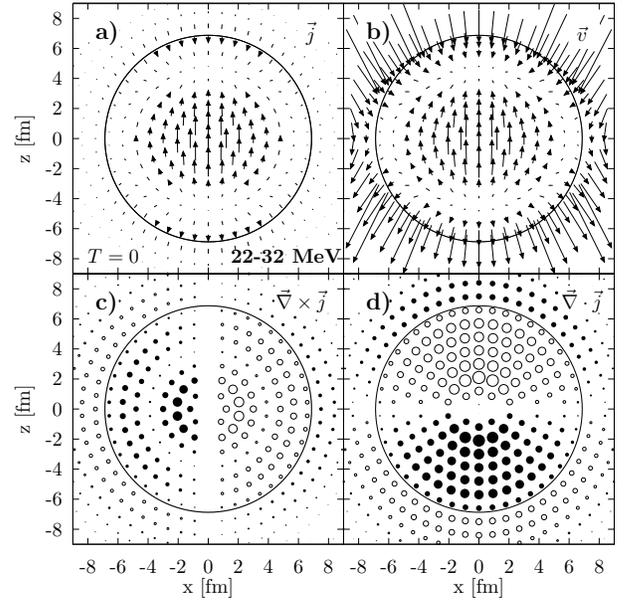}
\caption{E1(T=0) flow patterns in $^{208}$Pb for the
CM energy bin 22-32 MeV. See Fig. 4 for notation.}
\end{figure}
\begin{figure*}
\includegraphics[width=7cm,angle=-90]{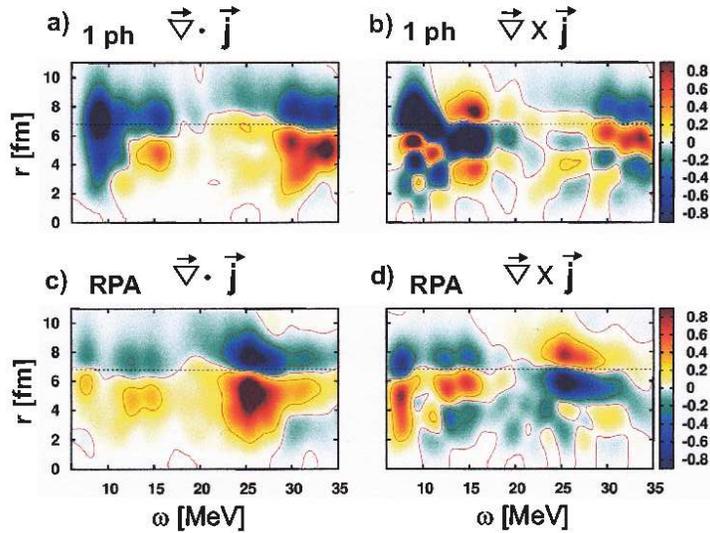}
\caption{ \label{fig:fig8}
 The coordinate-energy maps for the divergence (a,c) and curl
(b,d) of the nuclear current in E1(T=0) channel in $^{208}$Pb. The
upper (a,b) and lower (c,d) panels represent the 1ph and RPA
strengths, respectively. The nuclear radius is marked by the
dotted line. See explicit expressions for the patterns in Appendix
B.}
\end{figure*}
\begin{figure*}
\includegraphics[width=8cm,angle=-90]{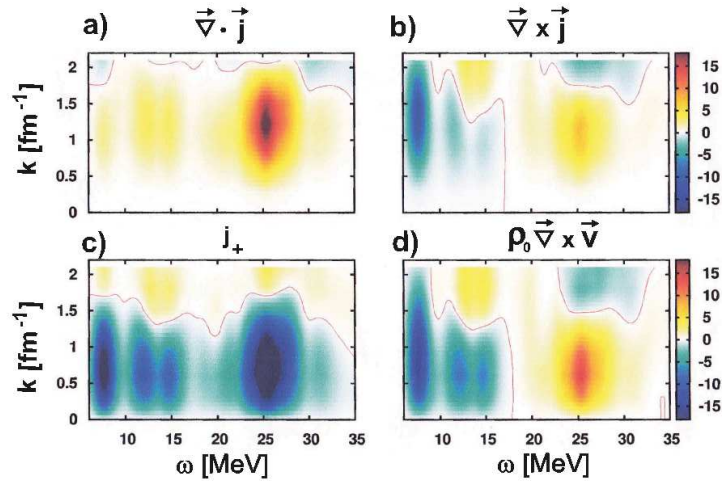}
\caption{\label{fig:fig9}
 The E1(T=0) RPA formfactors (\ref{ff}) in $^{208}$Pb for
 divergence of current (a), curl of current (b),
 $j_+$ component of the current (c), density-weighted velocity curl (d).
See explicit expressions for the patterns in Appendix B.}
\end{figure*}

As shown in Sec. 2, the TM and CM operators are composed from
$\vec j_+$ and $\vec j_-$ components of the nuclear current.
Moreover, following \cite{Ra87}, the component $\vec j_+$ is
treated as a measure of the nuclear vorticity. So it is worth to
inspect the $\vec j_+$ and $\vec j_-$ flows in more detail. The
relevant flow patterns are given in Fig. 5. It is seen (panels
(a)-(b)) that $\vec j_+$ and $\vec j_-$ are essentially different:
the former is maximal in the north and south poles while the later
is maximal in the nuclear center. Despite this difference, curls
of $\vec j_+$ and $\vec j_-$ are rather similar (panels (c)-(d)).
The same takes place (up to the total sign) for the divergencies
(panels (e)-(f)). Moreover, divergences and curls of $\vec j_+$
and $\vec j_-$ are of the same order of magnitude. So neither of
these current components alone is suitable to represent neither
vortical nor irrotational flows. Only their proper combinations,
like TM and CM ones, may be appropriate for this aim. The value
$\vec j_+$ has no any significant advantage over $\vec j_-$ as a
measure of the vorticity, which makes the RW vorticity criterium
\cite{Ra87} indeed questionable.

In Fig. 6, both fields $\vec j$ and $\vec v$ reproduce the typical
compression dipole motion (compare to Fig 1(b) for the schematic
image of CM). The divergence of the current is stronger than its
curl. This is natural for CM which, being almost irrotational, is
not, at the same time, the Tassie divergence-free mode.

\subsection{Coordinate-energy maps and form-factors}

In Fig. 7, the  smoothed coordinate-energy maps (\ref{aver_tcd})
for divergence and curl of the nuclear current are given for 1ph
(unperturbed Hartree-Fock)  and RPA E1(T=0) excitations. It is
seen that both $\vec{\nabla}\cdot \vec{j}$ and $\vec{\nabla}\times
\vec{j}$ are strong in a wide radial region 3 fm $<r<$ 10 fm
around the nuclear surface at $\sim$ 7 fm. Following panels (a,b),
the 1ph strength is concentrated in broad energy intervals:
low-energy (LE) 4-17 MeV and high-energy (HE)  28-35 MeV. In both
intervals, the curl and divergence are strong. The strength is
multi-modal, which is common for non-collective (single-particle)
excitations.

As seen from the panels (c,d) for the RPA case, inclusion of the
residual interaction considerably changes the pictures. Being
isoscalar, the residual interaction downshifts by energy both
$\vec{\nabla}\cdot \vec{j}$ and $\vec{\nabla}\times \vec{j}$. In
the CM region, the maxima are shifted from 30-35 MeV to 24-28 MeV.
The RPA distributions correspond to the strength functions
exhibited in Fig.~2 with the TM at $\sim$ 7 MeV, increased
vorticity at 12-15 MeV and 25-30 MeV, and irrotational CM at
$\sim$ 25 MeV.
\begin{figure*}
\includegraphics[width=4cm,angle=-90]{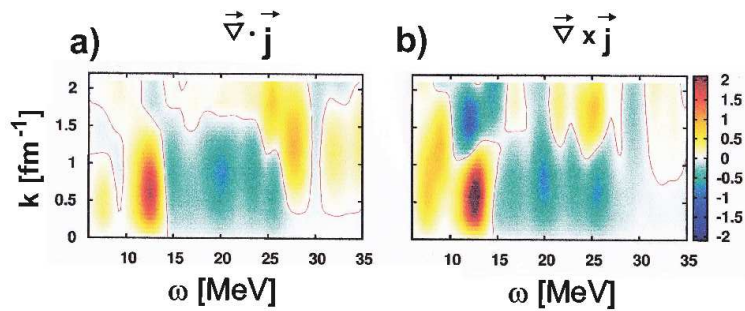}
\caption{ \label{fig:fig10} The E1(T=1) RPA formfactors (\ref{ff})
in $^{208}$Pb for divergence (a) and curl (b) of the current. See
explicit expressions for the patterns in Appendix B.}
\end{figure*}

It is remarkable that, after switch to RPA, both
$\vec{\nabla}\cdot \vec{j}$ and $\vec{\nabla}\times \vec{j}$
become weaker in the GDR region 10-15 MeV. For the first glance,
this result looks surprising. However both isovector GDR and
isoscalar spurious c.m. motion are collective Tassie modes for
which $\vec{\nabla}\cdot \vec{j}=\vec{\nabla}\times \vec{j}=0$.
Then the panels  (c)-(d) actually  show the rest of
$\vec{\nabla}\cdot \vec{j}$ and $\vec{\nabla}\times \vec{j}$ not
yet washed out by the dominant Tassie collective dipole motion. So
the Tassie motion can significantly suppress $\vec{\nabla}\cdot
\vec{j}$ and $\vec{\nabla}\times \vec{j}$ initially produced by
the single-particle motion. Instead, the CM and TM are not Tassie
modes and thus survive in the RPA case. The plots (c,d) show that
CM determined by $\vec{\nabla}\cdot \vec{j}$ (see
Eq.~(\ref{CM_divj})) is concentrated at $\sim$ 25 MeV while TM
determined by $\vec{\nabla}\times \vec{j}$ (see Eq.
(\ref{TM_curlj})) is distinctive at $\sim$ 7 MeV. Some
$\vec{\nabla}\times \vec{j}$ strength still remains at 12-15 MeV.

In Fig. 8, the smoothed E1(T=0) form-factors (\ref{ff}) for the
values of interest are presented. Namely,the values
$\vec{\nabla}\cdot \vec{j}$ , $\vec{\nabla}\times \vec{j}$,
current component $j_+$, and $\rho_0 \; \vec{\nabla}\times
\vec{v}$, pertinent for CM,TM, RW, and HD strengths, are
considered. Unlike the above transition coordinate-energy maps,
the form-factors are direct constituents of $(e,e')$ cross-section
and their inspection may suggest the most optimal transfer momenta
$k$ to observe a desirable mode. As follows from Fig. 9, the
observation of $\vec{\nabla}\cdot \vec{j}$  and
$\vec{\nabla}\times \vec{j}$, and thus related CM ($\sim$ 25 MeV)
and TM ($\sim$ 7 MeV), requests rather large momenta, 0.8 fm$^{-1}
< k <$ 1.6 fm$^{-1}$, which testifies that CM and TM are mainly
concentrated in the nuclear interior (which confirmed by Figs. 1,
4(a) and 7(a)). Instead the form-factors for $j_+$, and $\rho_0
\vec{\nabla}\times \vec{v}$ are maximal for lower momenta, 0.6
fm$^{-1} < k <$ 1.1 fm$^{-1}$, which points to their more surface
character. Note that $j_+$ has strict maxima in both low-energy TM
and high-energy CM regions. The form-factors for $\rho_0
\vec{\nabla}\times \vec{v}$ and $\vec{\nabla}\times \vec{j}$ are
similar, though the former is a bit stronger and shifted to lower
$k$. The difference at low $k$ arises  because these two
form-factors are mainly distinguished by the coordinate dependence
of the density $\rho_0(r)$, which is maximal at the nuclear
surface (= low $k$).

For the comparison, in Fig. 9, the isovector E1(T=1) RPA
form-factors for $\vec{\nabla}\cdot \vec{j}$  and
$\vec{\nabla}\times \vec{j}$ are depicted. It is seen that they
are weaker than in T=0 channel. The reason is again in the
presence of the dominant collective Tassie mode. Indeed, within
the Goldhaber-Teller model \cite{GT_48}, the GDR is essentially
the Tassie mode. Hence we have the strong suppression of
$\vec{\nabla}\cdot \vec{j}$ and $\vec{\nabla}\times \vec{j}$.
Nevertheless, in Fig. 10, the GDR region  still has noticeable
$\vec{\nabla}\cdot \vec{j}$ and $\vec{\nabla}\times \vec{j}$ at
12-13 MeV.  This could signal that the actual GDR is a combination
of the Goldhaber-Teller \cite{GT_48} (Tassie mode) and
Steinwedel-Jensen  \cite{SJ_50} (beyond Tassie mode) flows. There
are hints of the isovector TM at 11-13 MeV. The isovector CM is
not seen. Perhaps it is shifted above the energy 35 MeV (as
compared to the unperturbed 1ph CM strength at 29-35 MeV, depicted
in Fig. 8(a)). Comparison of Figs. 9 and 10 shows that the T=0
channel is more suitable for the experimental search of TM and CM
than the T=1 one.

\section{Discussion}


The strength functions, flow patterns, coordinate-energy maps, and
formfactors exhibited above show that $j_+$ component of the
current has no any essential advantage over $j_-$ as the vorticity
indicator. Indeed both components: i) are peaked in TM (basically
vortical) and CM (basically irrotational) regions, ii) have  curls
and divergences  of the same order of magnitude in TM region. This
indicates that $j_+$ or $j_-$ alone cannot be a relevant measure
of the vorticity. However, such a measure can be designed as a
proper combination of $j_+$ or $j_-$. The toroidal mode is a
natural case of such design. This transversal mode is free from
the longitudinal part arising in the long-wave approximation (LWA)
and its flow has a clear curl-like character.

As shown above, implementation of HD characteristics, like $\delta
{\vec v}$, is not convenient because of their unphysical behavior
at the nuclear surface and beyond. To demonstrate the HD
vorticity, it is better to use the toroidal flow which gives  a
similar vorticity and  is well behaved near the nuclear surface.
Altogether, the numerical arguments favor TM as a measure of the
vorticity.

Before discussing different aspects of nuclear vorticity, it is
worth to define criteria for the {\it vortical} nuclear current.
These could be: i) rotational flow pattern closely corresponding
to the HD view, ii) decoupling from the CE i.e. transversal
(divergence-free) character of the current. Such vortical nuclear
current should correctly manifest itself in the basic test cases
of TM and CM in E1(T=0) channel. Namely it has to dominate in the
TM which is mainly vortical  and vanish in the CM whose flow
pattern is mainly irrotational.

The above requirement ii) is closely related to the definition of
the independent current component (ICC) \cite{Ra87,Hei_82} which,
together with electric longitudinal (reduced to the nuclear
density) and magnetic transversal components, should constitute a
complete set describing the charge and current distributions in
the nucleus. There are at least two ways to define ICC.

The first way to determine ICC is proposed by Heisenberg
\cite{Hei_82} and later Rawenthall and Wambach \cite{Ra87}. Here
the decomposition of the nuclear current transition density
\begin{equation}
\delta \vec{j}(\vec{r}) = -i\sum_{\lambda \mu}
a_{\lambda \mu} \: (j_{\lambda  \lambda-1}(r) \:\vec{Y}^*_{\lambda
\lambda -1 \mu} + j_{\lambda  \lambda+1}(r) \:\vec{Y}^*_{\lambda
\lambda +1 \mu})
\end{equation}
in terms of  $j_-=j_{\lambda \lambda -1}(r)$ and $j_+= j_{\lambda
\lambda +1}(r)$ is used. The component $j_+$ is claimed
CE-unrestricted despite the CE (\ref{ce_ctd}) actually couples the
radial parts of $\delta \rho_{\lambda}$, $j_-$, and $j_+$:
\begin{eqnarray}\nonumber
 \omega \delta \rho_{\lambda}(r) =
 &-& \sqrt{\frac{\lambda}{2\lambda +1}}
 (\frac{d}{dr}-\frac{\lambda -1}{r}) j_{\lambda \lambda -1}(r)
 \\
 \label{CE_mm}
 &+& \sqrt{\frac{\lambda+1}{2\lambda +1}}
 (\frac{d}{dr}+\frac{\lambda +2}{r}) j_{\lambda \lambda +1}(r) \;
 .
\end{eqnarray}
The claim is based on the analysis of the multipole moments given
from the right and left sides of (\ref{CE_mm}). The moments for
$\delta \rho_{\lambda}$ and $j_-$ are coupled,
\begin{equation}
\omega \int dr r^{\lambda+2}\delta \rho_{\lambda}(r) =
\sqrt{\lambda(2\lambda +1)} \int dr
r^{\lambda+1}j_{\lambda\lambda-1}(r) \; ,
\end{equation}
while $j_+$-moments fully vanish. Therefore, $j_+$ is considered
as CE-unrestricted and thus suitable to represent ICC and nuclear
vorticity \cite{Ra87}. Following this prescription, vorticity of
the nuclear current is fully determined by its $j_+$-component.

By our opinion, this prescription is not good at least for the
following reasons. First, the vanishing of $j_+$-moments decouples
$j_+$ from CE only in the integral sense while preserving the
local coupling (\ref{CE_mm}). In other words, $\vec{\nabla}\cdot
\vec{j}_+ (\vec{r})$ is not locally zero. Indeed in Fig. 5e)  the
field $\vec{\nabla}\cdot \vec{j}_+ (\vec{r})$ is locally strong.
However it has different sign at $z>0$ and $z<0$ and thus can
vanish being integrated. Second,  following (\ref{CM_oper}), $j_+$
contributes to CM, which suggests a considerable vortical fraction
in CM flow. At the same time, we know that CM is basically
irrotational and has the gradient-like velocity \cite{Kva_PRC_11}.
Third, our numerical analysis of TM/CM strengths and
curl/divergences of the current does not reveal any essential
advantage of $j_+$ over $j_-$ as the vorticity indicator.
Altogether, the ansatz \cite{Ra87} to use $j_+$ as a measure of
the vorticity and ICC looks doubtful.

Another (and more natural) way to define ICC has been proposed by
Dubovik et al \cite{Dub_75_83}. Here the electric current
transition density is decomposed into the longitudinal and
transversal components,
\begin{eqnarray}
  \delta \vec{j}(\vec{r})&=&\delta \vec{j}_{\parallel}(\vec{r})
   + \delta \vec{j}_{\perp}(\vec{r}) \:,
   \\
\delta \vec{j}_{\parallel}(\vec{r}) &=& \vec{\nabla}
  \phi(\vec{r}), \quad
  \delta \vec{j}_{\perp}(\vec{r}) = \vec{\nabla}
  \times \vec{\nabla} \times (\vec{r}\chi(\vec{r}))
\end{eqnarray}
where $\phi(\vec{r})$ and $\chi(\vec{r})$ are some scalar
functions. As compared to the prescription \cite{Ra87,Hei_82},
this way looks more logical for the search of CE-unrestricted
divergence-free ICC. Now we get  $\delta \vec{j}_{\perp}$ as the
natural ICC candidate from very beginning.

The current components can be expanded in the basis of
eigenfunctions $\vec{\it J}^{(\kappa)}_{\lambda\mu k}({\vec r})$
($\kappa$ = -, 0, +) of the vector Helmholtz equation (the similar
expansion is familiar for the vector-potential, see e.g.
\cite{EG_book}). Then the transversal component reads
\begin{equation}
\delta \vec{j}_{\perp}(\vec{r})=\sum_{\lambda \mu k}\vec{\it
J}^{(+)}_{\lambda\mu k}({\vec r}) \; m^{(+)}_{\lambda\mu} (k)
\end{equation}
where $m^{(+)}_{\lambda\mu} (k)$ are electric transversal
formfactors and integration by $k$ is assumed, In the LWA ($k \to
0$), the transversal component is reduced to the longitudinal one.
After subtraction of the LWA part from $\delta \vec{j}_{\perp}$,
we get at $k>0$ the toroidal current density.  The transversal
character of the toroidal current is also seen from
(\ref{tor_rel_1}) and (\ref{TM_curl_me}). Being independent from
$\delta \vec{j}_{\parallel}$  and thus decoupled from CE, the
toroidal current can be considered both as ICC \cite{Dub_75_83}
and relevant vortical part of the complete nuclear current. Unlike
the prescription \cite{Ra87,Hei_82}, this vortical current is
built from both $j_+$ and $j_-$ components, see e.g.
(\ref{TM_curl_me}). Its vorticity corresponds to HD one, see Sec.
II C.  Besides, the relevance of the TM current as a measure of
the vorticity is confirmed by our numerical analysis of flow
patterns. Altogether, our analysis shows that just TM and its
current are best representatives of the nuclear vorticity.

Finally note that for more detailed study of the nuclear
vorticity, it is desirable to go beyond RPA by taking into account
the coupling to complex configurations, see e,g, the relevant
extensions \cite{CoBo01,Sever_08,Li08,bianco12,Ry02,Sol92}. Note
that, for the proper treatment of anharmonic effects, inclusion
only of two-phonon configurations may not be enough. The impact of
higher configurations and exact record of the Pauli principle are
also necessary, see discussion \cite{Sol92,Ne90}. All these
factors make anharmonic models very complicated. Anyway, before
performing these involved investigations, a mere RPA exploration
is desirable and this is just our case.

\section{Conclusions}

The problem of nuclear vorticity in isoscalar E1 excitations
(toroidal and compression modes - TM and CM) was scrutinized
within the Skyrme RPA with the force SLy6. A representative set of
characteristics (strength functions, flow pattern for currents and
velocities, curls and divergences of the current and its
components, coordinate-energy maps and formactors) was inspected.
Analysis of curls $\vec{\nabla}\cdot \vec{j}$ and divergences
$\vec{\nabla}\times \vec{j}$ of the nuclear current, as direct
indicators of the vortical/irrotational flow and coupling to the
continuity equation (CE), was especially important. Note that the
isovector GDR and isoscalar spurious c.m. motion, being the Tassie
collective modes, do not contribute to $\vec{\nabla}\cdot \vec{j}$
and  $\vec{\nabla}\times \vec{j}$. Instead, the TM and CM do not
belong the Tassie modes and for them the  curls and divergences
become informative.

The numerical and analytical analysis shows that, unlike the
prescription \cite{Ra87,Hei_82}, the nuclear vorticity is better
described not by $j_+$ component of the nuclear current but by its
transversal toroidal part \cite{Dub_75_83} composed from both
$j_+$ and $j_-$ components. 
The toroidal motion is well decoupled from continuity equation,
closely corresponds to the hydrodynamical picture of the
vorticity, and  provides a reasonable treatment of
vortical/irrotational flow in toroidal  and compression mods in
E1(T=0) channel.

\section*{Acknowledgments}
The work was partly supported by the GSI-F+E-2010-12,
Heisenberg-Landau (Germany - BLTP JINR), and Votruba- Blokhintsev
(Czech Republic - BLTP JINR) grants. P.-G.R. is grateful for the
BMBF support under Contracts No. 06 DD 9052D and No. 06 ER 9063.
The support of the research plan MSM 0021620859 (Ministry of
Education of the Czech Republic) and the grant of Czech Science
Foundation (13- 07117S) are appreciated. We thank Prof. J. Wambach
for useful discussions.

\appendix
\section{Curls and divergencies}
\label{Sec:app1}

The curl and divergence of the current E1 transitions densities
read:
\begin{equation}\label{A1_rot_j}
\vec{\nabla} \times \delta \vec{j}^{\nu}_{1}(\vec{r}) = i [rot \;
j]^{\nu}(r) \; \vec{Y}^*_{1 1}
\end{equation}
where
\begin{equation}
\label{A2_rot_j}
[rot \; j]^{\nu}(r) = \sqrt{\frac{2}{3}}
\frac{d}{dr}j_{10}^{\nu}(r) +
\sqrt{\frac{1}{3}} [\frac{d}{dr}+\frac{3}{r}] j_{12}^{\nu} (r)
\end{equation}
and
\begin{equation}\label{div_j_Y}
\vec{\nabla} \cdot \delta \vec{j}^{\nu}_{1\mu}(\vec{r}) = i [div
\; j]^{\nu}(r) \; Y^*_{1}
\end{equation}
where
\begin{equation}\label{div_j}
[div \; j]^{\nu} (r)=\sqrt{\frac{1}{3}} \frac{d}{dr} j_{10}^{\nu} (r)
- \sqrt{\frac{2}{3}}
[\frac{d}{dr}+\frac{3}{r}] j_{12}^{\nu}(r) \; .
\end{equation}

The velocity transition density can be decomposed like the current one
(\ref{CTD}):
\begin{equation}\label{dv}
\delta \vec{v}^{\nu}_{1\mu}(\vec{r}) = [v_{10}^{\nu} (r)
\vec{Y}^*_{1 0 \mu}(\hat{\vec r}) + v_{12}^{\nu} (r) \vec{Y}^*_{1
2 \mu}(\hat{\vec r})] \; ,
\end{equation}
with
\begin{equation}\label{dv_10_11}
v_{10}^{\nu} (r)= \frac{j_{10}^{\nu} (r)}{\rho_0(r)} \; , \quad
v_{11}^{\nu} (r)= \frac{j_{11}^{\nu} (r)}{\rho_0(r)} \; .
\end{equation}
Then
\begin{eqnarray}\label{rot_v}
\vec{\nabla} \times \delta \vec{v}^{\nu}_{1\mu}(\vec{r}) &=& i
[rot \; v]^{\nu}(r) \; \vec{Y}^*_{1 1}
\; ,
\\
\label{div_v}
\vec{\nabla} \cdot \delta \vec{v}^{\nu}_{1\mu}(\vec{r}) &=&
 [div \; v]^{\nu}(r) \;
Y^*_{1}
\end{eqnarray}
with
\begin{equation}\label{rot_v_nu}
[rot \; v]^{\nu}(r)=\sqrt{\frac{2}{3}}
\frac{d}{dr}v_{10}^{\nu}(r) +
\sqrt{\frac{1}{3}} [\frac{d}{dr}+\frac{3}{r}] v_{12}^{\nu} (r) \; ,
\end{equation}
\begin{equation}
\label{div_v_nu}
[div \; v]^{\nu} (r)=\sqrt{\frac{1}{3}} \frac{d}{dr} v_{10}^{\nu} (r)
- \sqrt{\frac{2}{3}}
[\frac{d}{dr}+\frac{3}{r}] v_{12}^{\nu}(r) \; .
\end{equation}

\section{Integral and average characteristics}
\label{Sec:app2}

The flows in Figs. 4-7 represent the integral vector variables (\ref{int_tcd_A}) in
$\{x,y=0,z\}$ cartesian plane, i.e.
$\vec{A}(\vec{r})=A_x(x,y=0,z)\vec{e}_x +A_z(x,y=0,z)\vec{e}_z$.
Namely, we use:
\begin{eqnarray}
\label{B1}
\vec{j} \to \vec{A}^{\nu}(\vec{r})&=& \delta\vec{j}^{\nu}(\vec{r}),\\
\label{B2} \vec{j}_{10} \to \vec{A}^{\nu}(\vec{r})&=&
j_{10}^{\nu}(r)\vec{Y}^*_{10}
,\\
\label{B3} \vec{j}_{12} \to \vec{A}^{\nu}(\vec{r})&=&
j_{12}^{\nu}(r)\vec{Y}^*_{12} ,
\end{eqnarray}

\begin{eqnarray}
\label{B4}
\vec{\nabla}\times \vec{j} \to \vec{A}^{\nu}(\vec{r})&=& [rot \; j]^{\nu}(r)
\vec{Y}^*_{11} ,\\
\label{B5}
\vec{\nabla}\times \vec{j}_{10} \to\vec{A}^{\nu}(\vec{r})&=&
  \sqrt{\frac{2}{3}}\frac{d}{dr}j_{10}^{\nu}(r)
  \vec{Y}_{11} , \\
\label{B6}
\vec{\nabla}\times \vec{j}_{12} \to\vec{A}^{\nu}(\vec{r})&=&
\sqrt{\frac{1}{3}}[\frac{d}{dr}+\frac{3}{r}]j_{12}^{\nu}(r)
  \vec{Y}^*_{11} ,
\\
\label{B7}
\vec{v} \to \vec{A}^{\nu}(\vec{r})&=& \delta\vec{v}^{\nu}(\vec{r}),\\
\label{B8} \rho_0(r) \vec{\nabla}\times \vec{v} \to
\vec{A}^{\nu}(\vec{r})&=& \rho_0(r) [rot \; v]^{\nu}(r)
\vec{Y}^*_{11} .
\end{eqnarray}
The values in (\ref{B1})-(\ref{B3}), (\ref{B4})-(\ref{B6}), and (\ref{B7},\ref{B8})
are taken from expressions (\ref{CTD}), (\ref{A1_rot_j},\ref{A2_rot_j}),
and (\ref{dv}-\ref{rot_v},\ref{rot_v_nu}), respectively.

The scalar divergences in Figs. 4-7 use the values
\begin{eqnarray}
\vec{\nabla}\cdot \vec{j} \to B^{\nu}(\vec{r})&=&
               [div \; j]^{\nu}(r) Y^*_1 , \\
\vec{\nabla}\cdot \vec{j}_{10} \to B^{\nu}(\vec{r})&=&
               \sqrt{\frac{1}{3}}\frac{d}{dr}j_{10}^{\nu}(r) Y^*_1 ,
\\
\vec{\nabla}\cdot \vec{j}_{12} \to B^{\nu}(\vec{r})&=&
                -\sqrt{\frac{2}{3}}[\frac{d}{dr}
                +\frac{3}{r}] j_{12}^{\nu}(r) Y^*_1 ,\\
\rho_0(r) \vec{\nabla}\cdot \vec{v} \to  B^{\nu}(\vec{r})&=&
\rho_0(r) [div \; v]^{\nu}(r) Y^*_1
\end{eqnarray}
from expressions (\ref{div_j_Y},\ref{div_j}) and
(\ref{div_v},\ref{dv_10_11},\ref{div_v_nu}). The divergences are
depicted in the figures as circles of the area proportional to
$B(x,y=0,z)$. The filled (open) circles mean the positive
(negative) sign of the variable.

Further, Figs.~8-10 give the average radial-energy maps
(\ref{aver_tcd}) and form-factors (\ref{ff}) for the values
\begin{eqnarray}
\vec{\nabla}\times \vec{j} \to C^{\nu}(r)&=&[rot \; j]^{\nu}(r) \;,\\
\vec{\nabla}\cdot \vec{j} \to C^{\nu}(r)&=&[div \; j]^{\nu}(r) \; ,
\end{eqnarray}
taken from expressions (\ref{A2_rot_j}) and (\ref{div_j}).

\end{document}